\documentclass[a4paper]{article}
\usepackage[margin=30mm]{geometry}
\usepackage{graphicx}
\usepackage{complexity}
\usepackage{url}
\usepackage{authblk}

\newcounter{one}
\newcounter{two}
\newcounter{three}
\newcounter{four}
\newcounter{five}
\newcommand{\ini}{\mathsf{s}}
\newcommand{\tar}{\mathsf{t}}

\newcommand{\X}{\mathcal{X}}
\newcommand{\Z}{\mathcal{Z}}
\newcommand{\sets}{\mathcal{S}}
\newcommand{\zind}{\mathcal{Z}_{\mathrm{ind}}}
\newcommand{\zsol}{\mathcal{Z}_{\mathrm{sol}}}
\newcommand{\labeln}{\mathsf{label}}
\newcommand{\rootn}{\mathsf{root}}
\newcommand{\child}[1]{\mathsf{child}_{#1}}
\newcommand{\remove}{\mathsf{remove}}
\newcommand{\add}{\mathsf{add}}
\newcommand{\swap}{\mathsf{swap}}

\usepackage[ruled,vlined,linesnumbered]{algorithm2e}
\usepackage{amsmath}
\usepackage{amssymb}

\usepackage{CJKutf8}

\usepackage{color}

\newcommand{\KJ}[2]{#2}

\newcommand{\TI}[2]{#2}

\newcommand{\reconfrule}{\mathcal{R}} 

\begin{document}
\title{ZDD-Based Algorithmic Framework for Solving Shortest Reconfiguration Problems} \author[1]{Takehiro Ito} \author[2]{Jun Kawahara} \author[3]{Yu Nakahata} \author[4]{Takehide Soh} \author[1]{Akira Suzuki} \author[5]{Junichi Teruyama}
\author[6]{Takahisa Toda} \affil[1]{Graduate School of Information Sciences,
Tohoku University, Sendai, Japan}
\affil[2]{Graduate School of Informatics,
Kyoto University, Kyoto, Japan}
\affil[3]{Graduate School of Science and Technology,
Nara Institute of Science and Technology, Ikoma, Japan}
\affil[4]{Information Infrastructure and Digital Transformation Initiatives
Headquarters, Kobe University, Kobe, Japan}
\affil[5]{Graduate School of Information Sciences,
University of Hyogo, Kobe, Japan}
\affil[6]{Graduate School of Informatics and Engineering,
The University of Electro-Communications, Chofu, Japan}
\date{}

\maketitle              
\begin{abstract}
This paper proposes an algorithmic framework for various reconfiguration problems using zero-suppressed binary decision diagrams (ZDDs), a data structure for families of sets.
In general, a reconfiguration problem checks if there is a step-by-step transformation between two given feasible solutions (e.g., independent sets of an input graph) of a fixed search problem such that all intermediate results are also feasible and each step obeys a fixed reconfiguration rule (e.g., adding/removing a single vertex to/from an independent set).
The solution space formed by all feasible solutions can be exponential in the input size, and indeed many reconfiguration problems are known to be \PSPACE-complete.
This paper shows that an algorithm in the proposed framework efficiently conducts the breadth-first search by compressing the solution space using ZDDs, and finds a shortest transformation between two given feasible solutions if exists.
Moreover, the proposed framework provides rich information on the solution space, such as the connectivity of the solution space and all feasible solutions reachable from a specified one.
We demonstrate that the proposed framework can be applied to various reconfiguration problems, and experimentally evaluate their performances. 
\end{abstract}

\section{Introduction}

A combinatorial reconfiguration~\TI{}{\cite{IDHPSUU11,Nishimura18,Heuvel13}} is \TI{the problem}{a family of problems} of finding a
procedure that changes one solution of a combinatorial \TI{optimization}{search}
problem into another solution while \TI{satisfying}{maintaining} the
conditions of the \TI{}{search} problem, and has attracted much attention in
recent years. Taking the change of the switch configuration of a
power distribution network as an example, we can regard a switch
configuration \TI{on the network}{satisfying all the electric conditions} as a solution\TI{, and regard finding
the solution that minimizes the power distribution loss as a
combinatorial optimization problem}{~for a search problem~\cite{InoueTWKYKTMH14}.} 
In a reconfiguration version of \TI{the}{this search} problem, the task is to find \TI{the}{a} changing
procedure from the current switch configuration to \TI{the optimal}{another (more desirable)}
one while \TI{satisfying}{maintaining} required \TI{}{electric} conditions such as not causing
power outages. We can say, in general, that a combinatorial
reconfiguration problem models a situation in which we change the current
configuration into \TI{the other}{another} one without allowing the system to stop.

Combinatorial reconfiguration problems have been actively studied
in the theoretical algorithms community in recent years\TI{~\cite{Nishimura18}.}{. (See surveys~\cite{Nishimura18,Heuvel13}.)} In
particular, combinatorial reconfiguration problems related to
graphs, such as independent sets and graph coloring, have been
\TI{well studied}{studied well}. Many of these studies are mainly from a theoretical
perspective, such as analysis of the computational complexity of
the problem \TI{depending on the properties of classes of the input graph, and}{with respect to graph classes;} 
there is little research on the applied aspects to the
best of the authors' knowledge.  
Since many reconfiguration
problems\TI{}{,} such as the independent set reconfiguration\TI{}{~\cite{KaminskiMM12}} and graph 4-coloring reconfiguration\TI{}{~\cite{BC09},} are \PSPACE-complete\TI{~\cite{BC09}}{}, it is hard to design an efficient algorithm. 
However, depending on
applications, the number of vertices in the input graph may be at
most tens or hundreds, and in such cases, we can expect the existence of an algorithm that works in an acceptable time.

There are various promising methods for solving combinatorial
optimization problems that appear in real applications, such as
integer programming, SAT solvers, genetic algorithms,
and metaheuristics. One approach that has attracted attention is the
use of zero-suppressed binary decision diagrams (ZDDs)~\cite{Minato1993,Castro2022}. A ZDD is
a data structure that compresses and compactly represents a
family of sets. By representing the solution set of a
combinatorial optimization problem as a ZDD and by performing set
operations of ZDDs, it is possible to find the optimal solution
by imposing constraints that integer programming methods and SAT
solvers are not good at treating.

\begin{figure}[t]
    \begin{center}
    \includegraphics[width=0.9\textwidth]{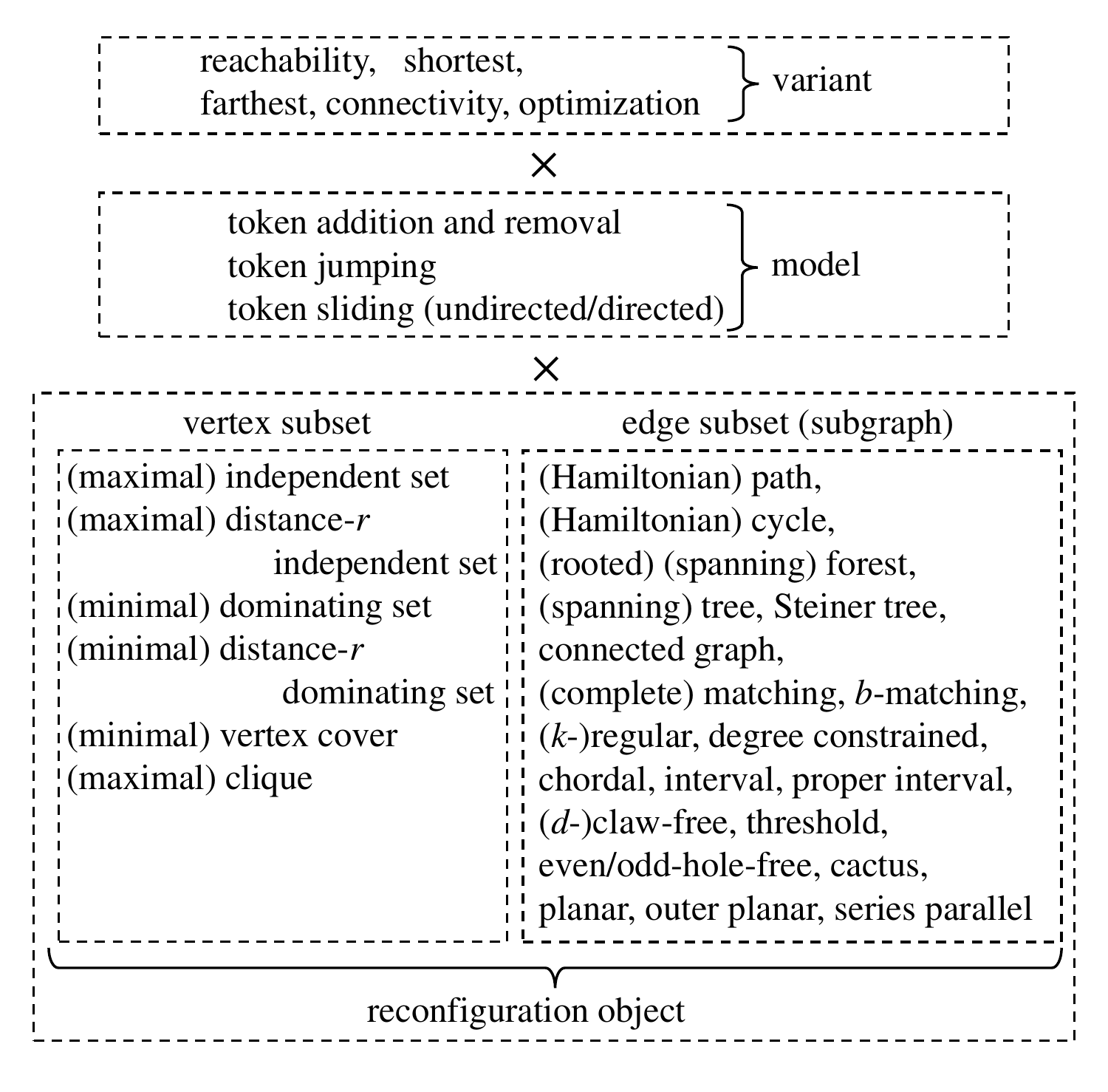}
    \end{center}
    \vspace{-2em}
    \caption{Variants, models, and reconfiguration \TI{problems}{objects} that our framework can treat.}
    \label{fig:problems}
\end{figure}

In this study, we consider solving combinatorial reconfiguration problems using ZDDs. 
We use the fact that a ZDD not only preserves one\TI{optimal}{} solution, but also \TI{all feasible}{\emph{all}} solutions. 
We propose an algorithm to obtain all solutions that are changeable from a given solution as a ZDD. 
\TI{Our}{This} algorithm can be applied to various combinatorial reconfiguration problems whose solution sets can be represented as ZDDs. 
\TI{We can also impose constraints that can be imposed on the solution set represented as ZDDs, such as the size of the solution. 
Our algorithm can solve the token jumping, token sliding, and token addition and removal models of the independent set reconfiguration problem (their definitions are introduced in the next section). 
\TI{Out}{Our} an algorithm can also handle reconfiguration not only for independent sets, but also for a variety of objects, such as dominating sets, vertex covers, spanning trees, and matchings, where we call them \emph{reconfiguration objects} (or simply \emph{objects}). Our algorithm can be used not only to analyze the changeability between two given solutions, but also to analyze the solution space, such as finding the shortest reconfiguration sequence, finding the solution with the longest shortest sequence from a given solution, deciding the connectivity of the solution space, and finding the optimal solution among solutions changeable from a given solution. What the proposed framework can treat is shown in Fig.~\ref{fig:problems}.}{Although we will give definitions in Section~\ref{sec:preliminaries}, \figurename~\ref{fig:problems} shows combinatorial reconfiguration problems that our algorithm can treat, 
where a combinatorial reconfiguration problem can be identified by the combination of problem variants, change rules (models), and solutions (reconfiguration objects). 
}
We demonstrate the effectiveness of the proposed algorithm by computer experiments.

The organization of the paper is as follows.
Section~\ref{sec:preliminaries} defines combinatorial reconfiguration problems and introduces ZDDs.
We propose an algorithm using ZDDs in Section~\ref{sec:framework}.
Section~\ref{sec:versatility} shows that the proposed algorithm can solve various combinatorial reconfiguration problems.
We conduct computer experiments in Section~\ref{sec:experiments} and conclude the paper in Section~\ref{sec:conclusion}.

\section{Preliminaries}\label{sec:preliminaries}

Throughout this paper, we use the symbols $G$, $V$, and $E$ to represent an input graph, its vertex set, and its edge set, respectively.
For families $\mathcal{A}$, $\mathcal{B}$ of sets,
we define $\mathcal{A} \Join \mathcal{B} =
\{A \cup B \mid A \in \mathcal{A}, B \in \mathcal{B} \}$.
In this paper, we sometimes simply call a family of sets a ``family.''

\subsection{Reconfiguration problems}\label{sec:reconf_problem}
\TI{}{As shown in \figurename~\ref{fig:problems}, a combinatorial reconfiguration problem can be identified by the combination of problem variants, models, and reconfiguration objects.

We first define reconfiguration objects.}
Throughout this paper, we use the symbol $U$ to represent a finite universal set, and assume that the \TI{objects for reconfiguration}{solutions for change} can be represented as subsets of $U$. 
More specifically, \KJ{given a reconfiguration problem,} we fix a property $\pi$ defined on the subsets of $U$, and say that a set $X \subseteq U$ is \TI{\emph{feasible}}{a \emph{reconfiguration object} or simply an \emph{object}} if $X$ satisfies $\pi$.

We \TI{also fix a}{then define models, also known as reconfiguration rules. A} \emph{reconfiguration rule} $\reconfrule$ on the subsets of $U$\TI{, which}{} defines whether two subsets of $U$ are \emph{adjacent} or not. 
\TI{}{There are three reconfiguration rules (called Token Addition and Removal, Token Jumping, and Token Sliding models) that have been studied well~\cite{KaminskiMM12,Nishimura18}, where we imagine that a \emph{token} is placed on each element in a subset of $U$. 
\smallskip

\noindent
\textbf{Token Addition and Removal:}
Two subsets $X$ and $Y$ of $U$ are adjacent under the \emph{token addition and removal} (TAR) model if and only if $|(X \setminus Y) \cup (Y \setminus X)| = 1$.
In other words, $Y$ can be obtained from $X$ by either adding a single element in $U \setminus X$ or removing a single element in $X$. 
\smallskip

\noindent
\textbf{Token Jumping:}
Two subsets $X$ and $Y$ of $U$ are adjacent under the \emph{token jumping} model if and only if $|X \setminus Y| = |Y \setminus X| = 1$.
Namely, $Y$ can be obtained from $X$ by exchanging a single element in $X$ with an element in $U \setminus X$. 
\smallskip

\noindent
\textbf{Token Sliding:}
This model is defined only on \KJ{graphs}{an input graph} $G = (V,E)$.
First, consider the case where $U = V$. 
Then, two subsets $X$ and $Y$ of $V$ are adjacent under the \emph{token sliding} model if and only if $X \setminus Y = \{u\}$, $Y \setminus X = \{v\}$, and $\{u, v\} \in E$.
Namely, $Y$ can be obtained from $X$ by exchanging a single vertex $u \in X$ with its \KJ{neighbor} adjacent vertex $v \in V \setminus X$. 
Next, consider the case where $U = E$. 
Then, two subsets $X$ and $Y$ of $E$ are adjacent under the \emph{token sliding} model if and only if $X \setminus Y = \{e\}$, $Y \setminus X = \{e'\}$ and $e$ and $e'$ share one of their endpoints. 
Namely, $Y$ can be obtained from $X$ by exchanging a single edge $e \in X$ with its incident edge $e' \in E \setminus X$. 
\smallskip

We finally define problem variants.
For} a given universal set $U$, the \emph{solution space} under the property $\pi$ and the reconfiguration rule $\reconfrule$ is a graph where each node corresponds to a reconfiguration object of $U$, and two nodes are joined by an edge if and only if their corresponding sets are adjacent under $\reconfrule$. 
Then, we can consider various variants of reconfiguration problems on the solution space: 
the \emph{reachability variant} asks whether the solution space contains a path connecting two given objects; 
the \emph{shortest variant} asks to compute the shortest length (i.e., the minimum number of edges) of any path in the solution space that connects two given objects;
the \emph{farthest variant} asks to find an object farthest from a given object in the solution space (i.e., the shortest path between the two objects is the longest);
the \emph{connectivity variant} asks whether the solution space is connected or not;
and the \emph{optimization variant} asks to find an object that is ``optimal'' (e.g., the cardinality of the object is maximum) among objects reachable from a given object.

\subsection{Zero-suppressed decision diagram (ZDD)}\label{sec:zdd}

A ZDD is a data structure for efficiently representing a family of sets.
The definition of a ZDD is given below.
In this section, we set $U = \{x_1, \ldots,x_n\}$
and $x_1 < x_2 < \cdots < x_n$.
A ZDD is a directed acyclic graph (DAG) $\Z$ that has the following properties:
$\Z$ has at most two nodes with outdegree zero, which are called \emph{terminals} and denoted by $\bot$ and $\top$.
Nodes other than the terminals are called non-terminal nodes.
A non-terminal node $\nu$ has an element in $U$, which is called a \emph{label} and
denoted by $\labeln(\nu)$,
and has two arcs, called \emph{0-arc} and \emph{1-arc}.
If the 0-arc and 1-arc of a non-terminal node $\nu$ point at nodes $\nu_0, \nu_1$,
we write $\nu = (\labeln(\nu), \nu_0, \nu_1)$, and we call $\nu_0$ and $\nu_1$ \emph{0-child}
and \emph{1-child}, respectively. Then, $\labeln(\nu) < \labeln(\nu_0)$
and $\labeln(\nu) < \labeln(\nu_1)$ must hold,
where we promise that $x_i < \labeln(\bot)$ and $x_i < \labeln(\top)$
for all $i = 1,\ldots,n$.
The ZDD $\Z$ has exactly one node with indegree zero, called the \emph{root}, and denoted by
$\rootn(\Z)$.

A ZDD $\Z$ represents a family of sets whose universal set is $U$ as follows.
We associate each node in $\Z$ with a family,
denoted by $\mathcal{S}(\nu)$, in the following recursive manner.
The terminal nodes $\bot$ and $\top$ are associated with
$\emptyset$ and $\{\emptyset\}$, respectively;
that is, $\mathcal{S}(\bot) = \emptyset$ and $\mathcal{S}(\top) = \{\emptyset\}$.
Consider the case where $\nu$ is a non-terminal node.
Let $\nu = (x, \nu_0, \nu_1)$, where $x \in U$ and $\nu_0, \nu_1$ are nodes of $\Z$.
The node $\nu$ is associated with the union of the family with which we associate $\nu_0$,
and the family obtained by adding $x$ to each set in
the family with which we associate $\nu_1$;
that is, $\mathcal{S}(\nu) = \mathcal{S}(\nu_0)
\cup (\{\{x\}\} \Join \mathcal{S}(\nu_1))$.
(Note that $\{\{x\}\} \Join \mathcal{S}(\nu_1)$ is the family obtained by adding $x$ to each set in $\mathcal{S}(\nu_1)$.)
Observe that each of the sets in $\mathcal{S}(\nu_0)$ and $\mathcal{S}(\nu_1)$
does not contain $x$ because of the property of ZDDs
i.e., $\labeln(\nu) < \labeln(\nu_0)$
and $\labeln(\nu) < \labeln(\nu_1)$ must hold.
We interpret that the ZDD $\Z$ represents the family with which we associate
the root node.
The family $\mathcal{S}(\Z)$ represented by the ZDD $\Z$
is defined by $\mathcal{S}(\Z) = \mathcal{S}(\rootn(\Z))$
(we use the same notation $\mathcal{S}$ for a node and a ZDD).

\begin{figure}[t]
  \centering
  \includegraphics[width=\textwidth]{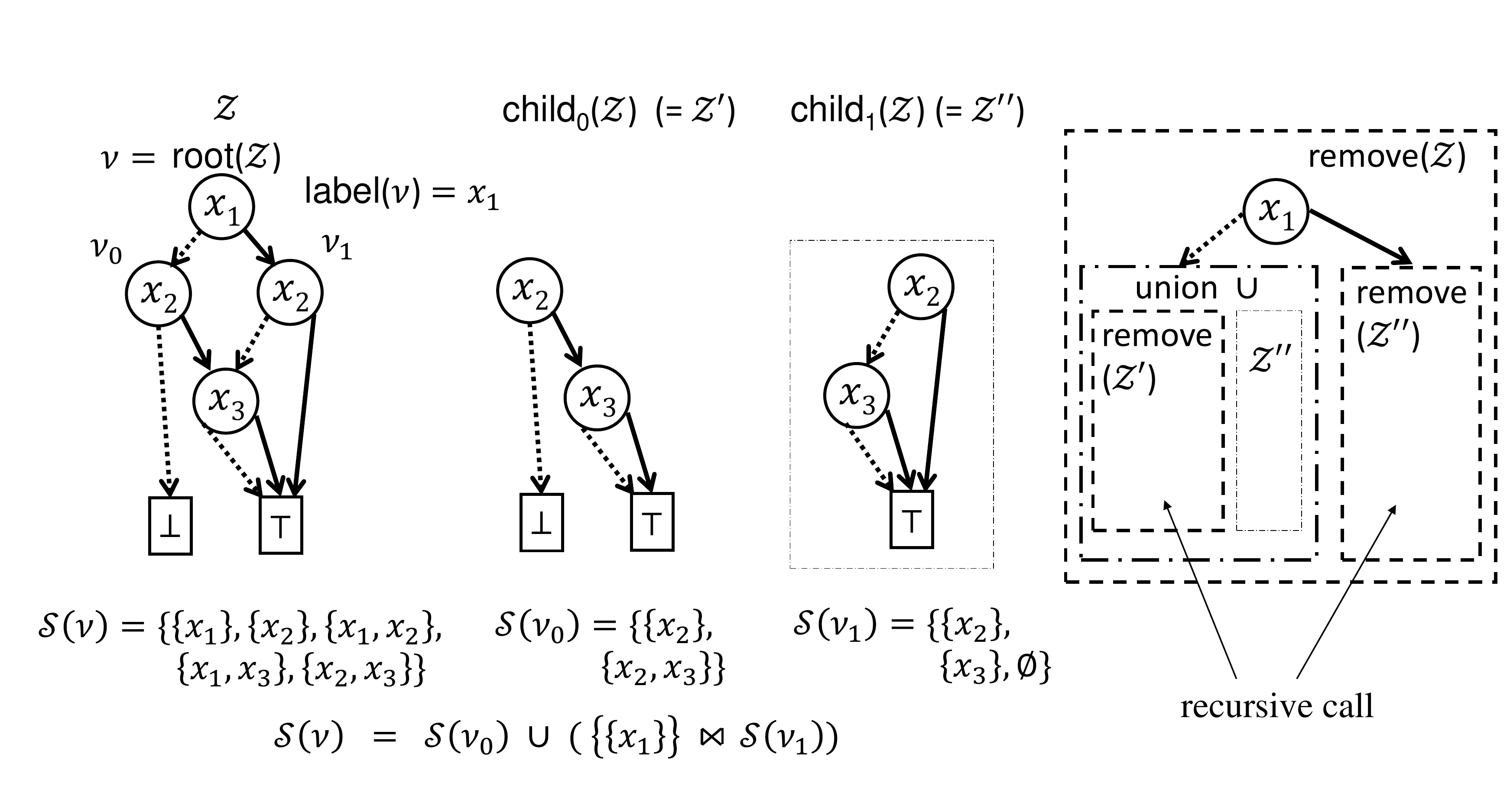}
  \caption{Example of a ZDD, the recursive structure, and the removal operation. The second argument of $\remove$ is omitted in the figure. Let $\Z'$ and $\Z''$ be $\child{0}(\Z)$ and $\child{1}(\Z)$, respectively. }
  \label{fig:zdd}
\end{figure}

Every ZDD $\Z$ has the following recursive structure~\cite{Bryant1986,Minato1993}.
Let $\nu = \rootn(\Z)$ and suppose that $\nu$ is represented by
$\nu = (x, \nu_0, \nu_1)$. Then, for $i = 0, 1$, the DAG consisting of the nodes and arcs
reachable from $\nu_i$ can be considered as a ZDD whose root is $\nu_i$, which we denote $\child{i}(\Z)$.
The ZDD $\child{i}(\Z)$ represents $\mathcal{S}(\nu_i)$.
To compute problems using a ZDD $\Z$, we often design a recursive algorithm $\mathsf{op}(\Z)$, which calls $\mathsf{op}(\child{0}(\Z))$ and $\mathsf{op}(\child{1}(\Z))$ and manipulates the results of them.
To make this clearer, we describe the behavior of the recursive algorithm for the set union operation, using \figurename~\ref{fig:zdd}.
Let us consider to construct the ZDD
 $\mathcal{Z}$ for the union of $\mathcal{S}=\{\{x_1\}, \{x_2\},\{x_1,x_2\}\}$ and $\mathcal{S}'=\{\{x_1,x_3\},\{x_2,x_3\}\}$ from the ZDDs of $S$ and $S'$.
At first, we focus on sets in $\mathcal{S}$ and $\mathcal{S}'$ that do not include $x_1$ and take union over them.
We do this by applying union operation for the left children of the ZDDs for $\mathcal{S}$ and for $\mathcal{S}'$.
The operation is done in a recursive manner: the base step is when one of ZDDs is $\top$ or $\bot$, which is straightforward and omitted.
The induction step is about to be described below.
The same applies to the union for the other sets, that is, the sets in $\mathcal{S}$ and $\mathcal{S}'$ that include $x_1$.
Here we have the two ZDDs as a result of applying recursive operation: one represents all sets in $\mathcal{S}\cup\mathcal{S}'$ that do not include $x_1$ and the other represents those in $\mathcal{S}'$ that include $x_1$, which correspond to $\mathcal{Z}_0$ and $\mathcal{Z}_1$ in \figurename~\ref{fig:zdd}, respectively.
We thus construct the final ZDD $\mathcal{Z}$ so that the $0$-arc of $x_1$ points to $\mathcal{Z}_0$ and the $1$-arc points to $\mathcal{Z}_1$.

Rich operations for manipulating families are provided for ZDDs~\cite{Bryant1986,Minato1993}.
For example, given two ZDDs $\Z, \Z'$, we can efficiently compute ZDDs representing
$\mathcal{S}(\Z) \cup \mathcal{S}(\Z')$, $\mathcal{S}(\Z) \cap \mathcal{S}(\Z')$,
$\mathcal{S}(\Z) \setminus \mathcal{S}(\Z')$, and so on,
using recursive ways described above.
For a binary operation $\circ \in \{ \cup, \cap, \setminus,\ldots\}$,
we denote the ZDD representing $\mathcal{S}(\Z) \circ \mathcal{S}(\Z')$
by $\Z \circ \Z'$.
For more information on ZDDs, please refer to \cite{knuth:taocp41}.

\section{ZDD-Based algorithmic framework}\label{sec:framework}

\subsection{Algorithmic framework}\label{sec:framework_overview}

We begin with the reachability variant under the TAR model for independent sets, where reconfiguration objects are independent sets in an input graph $G$ of cardinality at least a given threshold $k$.
Hayase et al.~\cite{Hayase1995} proposed an algorithm that constructs a ZDD, say $\zind$,
representing the family of all the independent sets
of a given graph, where the universal set $U$ is the vertex set of the graph and the elements (vertices) in $U$ are ordered.
The ZDD $\zind$ is highly compressed if a given graph
has a good structure, such as one with small pathwidth.
For example, an $8 \times 250$ grid graph has
$3.07 \times 10^{361}$ independent sets, but the ZDD representing them
has just $49{,}989$ nodes (with about 1MB memory usage).

We consider utilizing a ZDD that represents a vast number of independent sets for the reconfiguration problem.
Although $\zind$ includes all the independent sets,
it does not have information on the adjacency relations of independent sets.
What we would like to obtain is the family of independent sets adjacent to a given independent set
and more generally, the family of independent sets
adjacent to any of the independent sets in a given family.
If we can obtain them, by repeating the operation of obtaining the family of adjacent independent sets from the initial independent set,
we can obtain all the independent sets reachable from the initial independent set and decide whether a reconfiguration sequence from the initial independent set to the target one exists.

The TAR model requires two operations: removing a vertex from an independent set and adding a vertex to an independent set.
First, we consider the removal operation.
Given the family $\mathcal{I}$ of independent sets, the removal operation is to remove each element from each independent set in $\mathcal{I}$,
i.e., to obtain the family $\{ I \setminus \{ v\} \mid I \in \mathcal{I}, v \in I \}$.
Given $\mathcal{I}$ as a ZDD,
we propose an algorithm that constructs
a ZDD representing $\{ I \setminus \{ v\} \mid I \in \mathcal{I}, v \in I \}$ without extracting elements from the ZDD.
As the addition operation,
we also propose an algorithm that constructs
a ZDD representing $\{ I \cup \{ v\} \mid I \in \mathcal{I}, v \in U \setminus I \}$.

For later use, we present the two operations in a bit general form.
For a ZDD $\Z$ (whose universal set is $U$) and a set $R \subseteq U$,
let $\remove(\Z, R)$ be the ZDD representing
\[
    \{ I \setminus \{ v\}
        \mid I \in \mathcal{S}(\Z), v \in I \cap R \},
\]
which means that we remove an element only in $R$.
For a ZDD $\Z$ and a set $A \subseteq U$,
let $\add(\Z, A)$ be the ZDD representing
\[
    \{ I \cup \{x\} \mid I \in \mathcal{S}(\Z), x \in A \setminus I \},
\]
which means that we add an element only in $A$.

We can solve the reachability variant of the TAR model of the independent set reconfiguration problem using $\remove(\Z, R)$ and $\add(\Z, A)$ as follows.
First, we construct a ZDD representing the solution space.
Recall that a feasible independent set in the TAR model contains at least $k$ vertices.
It is easy to construct the ZDD, say $\mathcal{Z}_{\ge k}$, representing the family of all the sets with cardinality at least $k$ (i.e., $\{ I \subseteq U \mid |I| \ge k \}$)~\cite{knuth:taocp41}.
The solution space ZDD $\zsol$ is obtained by the intersection operation of $\zind$ and $\mathcal{Z}_{\ge k}$  mentioned in Sec.~\ref{sec:zdd}.

Next, for $i = 0,1,\ldots$,
let $\Z^i$ denote the ZDD representing the family of
independent sets obtained by applying
the reconfiguration rule (i.e., removing or adding a vertex) to $S$ exactly $i$ times,
where $\Z^0$ is the ZDD such that $\sets(\Z^0) = \{S\}$.
The construction of $\Z^0$ is trivial.
For $i = 1,2,\ldots$,
the ZDD $\Z^i$ can be constructed by
\begin{equation}\label{eq:z_i}
    \Z^i \gets  \mathsf{op}(\Z^{i-1}) \cap \zsol,
    \end{equation}
where $\mathsf{op}(\Z) =  \remove(\Z, V) \cup \add(\Z, V)$ for ZDD $\Z$.
Note that $\cup$ and $\cap$ are ZDD operations mentioned in Sec.~\ref{sec:zdd}.
After we construct $\Z^i$, we decide whether $\Z^i = \bot$ (i.e., $\mathcal{S}(\Z^i) = \emptyset$) and whether $T \in \mathcal{S}(\Z^i)$
(both are straightforward tasks).
If $\Z^i = \bot$, it means that the reconfiguration sequence from $S$ to $T$ does not exist because $\mathcal{S}(\Z^j)$, $0 \le j \le i$, includes \emph{all} the independent sets that are reachable from $S$ within $j$ steps.
We output NO and halt.
If $T \in \mathcal{S}(\Z^i)$, the reconfiguration sequence with length $i$ from $S$ to $T$ exists.
We output YES and halt.
If both do not hold, we next construct $\Z^{i+1}$.

\subsection{Removal and addition operations}

Given a ZDD $\Z$ and two sets $A$ and $R$,
we describe how to construct the ZDDs $\remove(\Z, R)$
and $\add(\Z, A)$.

We design an algorithm for $\remove(\Z, R)$ for a ZDD $\Z$, based on the recursive way described in Sec.~\ref{sec:zdd}. 
Let $\nu = \rootn(\Z)$. Suppose that $\nu$ is a non-terminal node and $\nu = (x, \nu_0, \nu_1)$, where $x \in U$ and $\nu_i$ is the $i$-child of $\nu$.

We consider the case of $x \in R$.
Let $\Z^{\mathrm{rem}} = \remove(\Z, R)$ and let us observe characteristics of $\Z^{\mathrm{rem}}$.
First, $\rootn(\Z^{\mathrm{rem}}) = x$ because $\sets(\Z^{\mathrm{rem}})$ contains a set including $x$ and does not contain any set including an element smaller than $x$.
Secondly, let us consider $\child{0}(\Z^{\mathrm{rem}})$,
which is the ZDD representing the family of sets in $\sets(\Z^{\mathrm{rem}})$ not containing $x$.
Each set in $\sets(\child{0}(\Z^{\mathrm{rem}}))$ is obtained by
either of the following two:
(i) we remove an element from a set in $\sets(\Z)$
not including $x$
(i.e., a set in $\sets(\child{0}(\Z))$), and
(ii) we remove $x$ from a set in $\sets(\Z)$ including $x$
(i.e., a set in $\{\{x\}\} \Join \sets(\child{1}(\Z))$).
We collect all the sets of (i) and construct the ZDD representing them.
The ZDD is obtained by recursively applying the $\remove$ operation to $\child{0}(\Z)$.
The ZDD for (ii) is just $\child{1}(\Z)$. Therefore, we obtain
\[
    \child{0}(\Z^{\mathrm{rem}}) = \remove(\child{0}(\Z), R \setminus \{x\}) \cup \child{1}(\Z),
\]
where `$\cup$' is the union operation of ZDDs described in Sec.~\ref{sec:zdd}.

Thirdly, we consider $\child{1}(\Z^{\mathrm{rem}})$,
which is the ZDD representing the family of sets
each of which is obtained by removing $x$
from a set in $\sets(\Z^{\mathrm{rem}})$ containing $x$.
Each set in $\sets(\child{1}(\Z^{\mathrm{rem}}))$ is obtained by
removing $x$ from a set in $\sets(\Z^{\mathrm{rem}})$ containing $x$.
The ZDD is obtained by applying the $\remove$ operation to $\child{1}(\Z)$. Therefore, we have
\[
    \child{1}(\Z^{\mathrm{rem}}) = \remove(\child{1}(\Z), R \setminus \{x\}).
\]

We consider the case of $x \notin R$, which means that we do not remove $x$ from any independent set.
Then, we obtain
\begin{align*}
    \child{0}(\Z^{\mathrm{rem}}) = \remove(\child{0}(\Z), R \setminus \{x\}),\\
    \child{1}(\Z^{\mathrm{rem}}) = \remove(\child{1}(\Z), R \setminus \{x\}).
\end{align*}

Our recursive algorithm for $\remove(\Z, U)$ is as follows:
If $\Z = \bot$ or $\Z = \top$, we just return $\bot$.
Otherwise, let $x = \labeln(\rootn(\Z))$, construct
\begin{align*}
    \Z_0 \gets & 
    \begin{cases}
    \remove(\child{0}(\Z), R \setminus \{x\}) \cup \child{1}(\Z) & \text{if } x \in R,\\
    \remove(\child{0}(\Z), R \setminus \{x\}) & \text{if } x \notin R,\\
    \end{cases}\\
    \Z_1 \gets & \remove(\child{1}(\Z), R \setminus \{x\}),
\end{align*}
and just call and return $\mathsf{makenode}(x, \Z_0, \Z_1)$ (see the right of Fig.~\ref{fig:zdd}),
where $\mathsf{makenode}(x, \Z_0, \Z_1)$ is the following procedure:
If there is a node whose label is $x$ and whose $i$-arc points at the root of $\Z_i$ for $i = 0, 1$, just return the node.
Otherwise, make a new node with label $x$, make its $i$-arc point at the root of $\Z_i$ for $i = 0, 1$, and return the new node.

Next, we design an algorithm for $\add(\Z, A)$ for any $A \subseteq U$.
Note that there is a possibility that an element that never appears in $\Z$ but in $A$ is added to a set.
Let $x = \labeln(\rootn(\Z))$ and $y$ be the minimum element in $A$.
First, we consider the case of $x \ge y$.
Similarly to the $\remove$ operation, we call
$\mathsf{makenode}(x, \Z_0, \Z_1)$,
where
\begin{align*}
    \Z_0 \gets & \add(\child{0}(\Z), A \setminus \{x\}),\\
    \Z_1 \gets & 
    \begin{cases}
    \add(\child{1}(\Z), A \setminus \{x\}) \cup \child{0}(\Z), & \text{if } x \in A \\
    \add(\child{1}(\Z), A \setminus \{x\}) & \text{if } x \notin A.
    \end{cases}
\end{align*}
We consider the case of $x < y$, including the case where $\Z = \top$ and $A \neq \emptyset$,
which means that the constructed family contains sets obtained by adding $y$ to sets in $\sets(\Z)$.
In this case, we consider a ZDD $\Z'$ equivalent to
$\Z$ (i.e., $\sets(\Z') = \sets(\Z)$) such that $\labeln(\rootn(\Z')) = y$.
Such $\Z'$ is constructed
by calling $\mathsf{makenode}(y, \Z, \bot)$.
We then call and return $\add(\Z', A)$ recursively.

At the end of the recursion, $\add(\bot, A) = \bot$ for any $A \subseteq U$
and $\add(\top, \emptyset) = \bot$ (the case where $\add(\top, A)$ for a non-emptyset $A$ has already been described above).

\section{Versatility of the proposed algorithm}\label{sec:versatility}

In this section, we show the versatility of the proposed algorithm in the following three directions:
(i) By using $\Z^i$ (in Sec.~\ref{sec:framework_overview}), we can solve the variants introduced in Sec.~\ref{sec:reconf_problem} (discussed in Sec.~\ref{sec:variants}); (ii) By changing $\mathsf{op}(\Z)$ in equation (\ref{eq:z_i}), we can solve some models (in Sec.~\ref{sec:models});
(iii) By constructing $\zsol$, we can treat various reconfiguration objects and constraints (in Sec.~\ref{sec:objects}).

\subsection{Shortest, farthest, connectivity, and optimization variants}\label{sec:variants}

The ZDD $\Z^i$ represents the family of \emph{all} the independent sets that are reachable from the initial set $S$ in $i$ steps.
Therefore, the smallest integer $i$ such that $T \in \sets(\Z^i)$ holds
is the length of a shortest reconfiguration sequence from $S$ to $T$.
The proposed algorithm can solve not only the reachability variant but also the shortest one.

The shortest sequence $I_0\ (= S),\ldots,I_h\ (= T)$ between $S$ and $T$ can be obtained by
the following backtrack method, where $h$ is the smallest integer such that
$T \in \sets(\Z^h)$.
Here, we consider only the token jumping model; other models are similar.
Suppose that we have already obtained $I_p,\ldots,I_h$ $(2 \le p \le h)$.
Then, there are vertices $v \notin I_p$ and $w \in I_p$
such that $I_p \cup \{v\} \setminus \{w\} \in \sets(\Z^{p-1})$
according to the construction of $\Z^{p}$.
Thus, we let $I_{p-1} := I_p \cup \{v\} \setminus \{w\}$.
By the above method, we obtain $I_1,\ldots,I_h$.
Finally, it is obvious that $|I_0 \setminus I_1| = |I_1 \setminus I_0| = 1$ holds according to the construction of $\Z^1$,
which shows that the sequence $I_0,\ldots,I_h$ is certainly the reconfiguration sequence between $S$ and $T$.
The computation time is as follows.
We can test whether $I_p \cup \{v\} \setminus \{w\}$ is in $\sets(\Z^{p-1})$ or not
by a ZDD operation in $\mathrm{O}(|V|)$ time.
The number of candidates of $I_{p-1}$ is $\mathrm{O}(|V|^2)$.
Therefore, the computation time of obtaining the shortest sequence after constructing the ZDDs $\Z^{0},\ldots,\Z^h$ is $\mathrm{O}(h|V|^3)$.

Let us consider the farthest variant.
We construct $\Z^0,\Z^1,\ldots$ without checking $T \in \sets(\Z^{i})$ in the algorithm
until $\Z^{i} = \bot$ holds.
Let $h'$ be the smallest integer such that $\Z^{h'} = \bot$.
Then, a set in $\sets(\Z^{h'-1})$ is a farthest independent set from $S$.

We solve the connectivity variant based on the following idea.
If the solution space (graph) is connected, all the independent sets are reachable
from any set $S$.
Therefore, we randomly choose $S$ from $\sets(\zsol)$ by a ZDD operation and construct $\Z^0,\Z^1,\ldots,\Z^{h'-1}$ in the same way as the farthest variant.
Then, by examining whether
$\zsol$ is equivalent to $\bigcup_{i = 0,\ldots,h'-1} \Z^i$ or not,
we obtain the answer.
Note that checking the equivalency of two given ZDDs can be done in $\mathrm{O}(1)$ time
in many ZDD manipulation systems.

We can solve the optimization variant as follows, where the criterion of
the optimality is arbitrary as long as the optimal solution can be
extracted from a ZDD in polynomial time (e.g., the number of elements in the solution and the total weight of the solution~\cite{knuth:taocp41}).
Given a ZDD $\Z$, for each $i$,
we extract an optimal solution $J_i \subseteq U$ from $\sets(\Z^i)$.
Then, we choose the optimal one among $J_0,J_1,\ldots$
(if there are two or more such solutions with the same size, we choose one with minimum index).
By running the algorithm for the shortest variant that obtains a reconfiguration sequence from $S$ to the chosen solution,
we obtain the desired reconfiguration sequence.

We conclude this subsection by pointing out that our algorithm can solve
the reconfiguration problem with multiple start sets $S_1,\ldots,S_s$
and goal sets $T_1,\ldots,T_t$, where the task is to decide
whether a reconfiguration sequence between $S_j$ and $T_{j'}$ exists
for some $j, j'$ or not. Just let $\Z^0$ be the ZDD for
$\{S_1,\ldots,S_s\}$ and change deciding whether $T \in \sets(\Z^{i})$
into $T_{j'} \in \sets(\Z^{i})$ for some $j'$.

\subsection{Token jumping and token sliding models}\label{sec:models}

We consider the token jumping and token sliding models
by designing $\mathsf{op}(\Z)$ in equation (\ref{eq:z_i}).

The swap operation is to remove a vertex and add another vertex from and to an independent set, respectively.
For a ZDD $\Z$ (whose universal set is $U$) and sets $A, R \subseteq U$,
let $\swap(\Z, A, R)$ be the ZDD representing
\[
    \{ I \cup \{v\} \setminus \{v'\}
        \mid I \in \mathcal{S}(\Z), v \in A \setminus I, v' \in I \cap R \},
\]
which means that we add an element in $A$ and remove an element in $R$.
This can be represented by
$\add(\remove(\Z^{i-1}, R), A) \setminus \Z^{i-1}$.
The set subtraction of $\Z^{i-1}$ is needed because the  family represented by $\add(\allowbreak \remove(\Z^{i-1}, R), A)$ includes sets obtained by removing and adding the same vertex.

We can design a more efficient algorithm.
We only show $\Z_0$ and $\Z_1$ when calling $\mathsf{makenode}(x, \Z_0, \Z_1)$ with $x = \rootn(\Z)$.
The others are similar to the addition operation.
$\Z_0$ and $\Z_1$ are
\begin{align*}
    \Z_0 \gets &
    \begin{cases}
    \swap(\child{0}(\Z), A \setminus \{x\}, R \setminus \{x\}) \cup \add(\child{1}(\Z), A \setminus \{x\}), & x \in R, \\
    \swap(\child{0}(\Z), A \setminus \{x\}, R \setminus \{x\}) & x \notin R,
    \end{cases}\\
    \Z_1 \gets & 
    \begin{cases}
    \swap(\child{1}(\Z), A \setminus \{x\}, R \setminus \{x\}) \cup \remove(\child{0}(\Z), R \setminus \{x\}), & x \in A, \\
    \swap(\child{1}(\Z), A \setminus \{x\}, R \setminus \{x\}) & x \notin A.
    \end{cases}
\end{align*}
This holds because $\sets(\Z_0)$ includes the independent sets 
obtained by removing $x$ from each set in $\{\{x\}\} \Join \sets(\child{1}(\Z))$ and then adding a vertex other than $x$ if $x \in R$,
and $\sets(\Z_1)$ includes the independent sets
obtained by adding $x$ to each set in $\sets(\child{0}(\Z))$
and removing a vertex other than $x$ if $x \in A$.
At the end of the recursion, $\swap(\bot, A, R) = \swap(\top, A, R) = \bot$ holds for any $A$ and $R$.

The slide operation can be performed similarly to the swap operation.
Let $\mathsf{slide}(\Z)$ be the ZDD representing
\[
    \{ I \cup \{v\} \setminus \{v'\}
        \mid I \in \mathcal{S}(\Z), v \in V \setminus I, v' \in I, \{v, v'\} \in E \}.
\]
We only show $\Z_0$ and $\Z_1$ when calling $\mathsf{makenode}(x, \Z_0, \Z_1)$ with $x = \rootn(\Z)$.
For a vertex $v \in V$, let $N(v)$ be the set of neighbors of $v$;
that is, $N(v) = \{ w \in V \mid \{v, w\} \in E\}$.
Then,
\begin{align}
    \Z_0 \gets & \mathsf{slide}(\child{0}(\Z)) \cup \add(\child{1}(\Z), N(x)), \label{eq:slide_add} \\
    \Z_1 \gets & \mathsf{slide}(\child{1}(\Z)) \cup \remove(\child{0}(\Z), N(x)). \label{eq:slide_remove}
\end{align}
The reason the second argument in $\add$ and $\remove$ in equations (\ref{eq:slide_add}) and (\ref{eq:slide_remove}) is $N(v)$
is that the $\add$ (resp. $\remove$) operation is called with $\child{1}(\Z)$ (resp. $\child{0}(\Z)$), which means that we remove (resp. add) $x$ from (resp. to) $\sets(\Z)$ and thus we must add (resp. remove) a vertex adjacent to $x$.

The algorithm can solve the token sliding model for a directed graph~\cite{ito_et_al:LIPIcs.MFCS.2022.58}.
The slide operation on a directed graph can be designed
with a slight modification:
\begin{align*}
    \Z_0 \gets & \mathsf{slide}(\child{0}(\Z)) \cup \add(\child{1}(\Z), N^{+}(x)), \\
    \Z_1 \gets & \mathsf{slide}(\child{1}(\Z)) \cup \remove(\child{0}(\Z), N^{-}(x)),
\end{align*}
where $N^{+}(v)$ is the heads of the arcs outgoing from $v$
and $N^{-}(v)$ is the tails of the arcs incoming to $v$ for $v \in V$.

\subsection{Reconfiguration objects and constraints}\label{sec:objects}

The proposed algorithm does not depend on the characteristics of independent sets except for the construction of $\zsol$.
Therefore, when we want to solve some reconfiguration problem,
we can use the proposed algorithm
by presenting how to construct $\zsol$ for objects corresponding to the problem.
Many researchers have proposed ZDD construction algorithms for various set families, some of which can be applied to ZDD construction for many reconfiguration objects.
In this section, we overview what kinds of objects we can treat.

\subsubsection{Vertex subset}

We show that we can construct many kinds of objects each of which is represented as a subset of the vertex set by set operations. Assume that the universal set $U$ is $V$.

We begin with independent sets (although we mentioned in the previous section that there is a more efficient algorithm~\cite{Hayase1995}).
Let $\X_v$ and $\overline{\X}_v$ be respectively the family of all the sets including $v$ and the family of those not including $v$;
that is, $\X_v = \{ A \subseteq U \mid v \in A \}$
and $\X_v = \{ A \subseteq U \mid v \notin A \}$.
It is easy to construct ZDDs for $\X_v$ and $\overline{\X}_v$.
For two vertices $v$ and $w$, the family of all the sets including
at most one of $v$ and $w$ is $\overline{\X}_v \cup \overline{\X}_w$.
Therefore, the family of all the independent sets is
\[
    \bigcap_{\{v, w\} \in E} \left( \overline{\X}_v \cup \overline{\X}_w \right),
\]
and the ZDD for this family can be simply obtained by combining known ZDD operations~\cite{knuth:taocp41}.
Similarly, we can solve other reconfiguration objects by ZDD operations~\cite{knuth:taocp41}.
We show some of them in Table~\ref{tab:zdd_vertex_subset}.

\begin{table}[t]
\caption{Reconfiguration objects that can be represented as vertex subsets, and how to obtain them by set operations~\cite{knuth:taocp41}. Let $N_k(v)$ be the set of vertices whose distance from $v$ ranges from $1$ to $k$.}
\label{tab:zdd_vertex_subset}
\centering
\begin{tabular}{|l|l|} \hline
Reconfiguration object & Set operations \\ \hline\hline
Dominating set & $\bigcap_{v \in V} \left( \X_v \cup \left(\bigcup_{w \in N(v)}\X_w\right) \right)$ \\ \hline
Vertex cover & $\bigcap_{\{v,w\} \in E} \left( \X_v \cup \X_w \right)$ \\ \hline
Clique & $\bigcap_{\{v, w\} \notin E} \left( \overline{\X}_v \cup \overline{\X}_w \right)$ \\ \hline
Distance-$k$ independent set & $\bigcap_{v \in V} \bigcap_{w \in N_k(v)} \left( \overline{\X}_v \cup \overline{\X}_w \right)$ \\ \hline
Distance-$k$ dominating set & $\bigcap_{v \in V} \left( \X_v \cup \left(\bigcup_{w \in N_k(v)}\X_w\right) \right)$ \\ \hline
\end{tabular}
\end{table}

Coudert~\cite{582363} proposed algorithms that construct ZDDs representing the families of sets
obtained by collecting only the maximal/minimal sets in a family given as a ZDD $\Z$;
that is, $\mathsf{maximal}(\Z)$ is the ZDD for $\{ X \in \sets(\Z) \mid \forall X' \in \sets(\Z), X \subseteq X' \implies X = X' \}$ and $\mathsf{minimal}(\Z)$ is the ZDD for $\{ X \in \sets(\Z) \mid \forall X' \in \sets(\Z), X' \subseteq X \implies X = X' \}$.
Using the maximal operation, we can solve (the token jumping model of) the maximal independent set reconfiguration problem~\cite{CENSORHILLEL202085}, where every feasible solution of this problem is a maximal independent set.
We can also solve some maximal/minimal reconfiguration problems, such as
the minimal dominating set reconfiguration, the minimal vertex cover reconfiguration, and the maximal clique reconfiguration.

\subsubsection{Subgraphs}

We consider a subgraph that can be represented by an edge set.
For example, a path can be represented by the set of edges consisting of the path.
Formally, for an edge set $E' \subseteq E$, a subgraph is represented by $(V', E')$,
where $V' = \bigcup_{\{v, w\} \in E'} (\{v\} \cup \{w\})$.
Then, we set the universe set $U$ to be $E$.
Note that we cannot treat subgraphs including isolated vertices in this representation.

Sekine et al.~\cite{Sekine1995} proposed an algorithm that constructs the ZDD
representing the family of all the spanning trees.
Knuth~\cite{knuth:taocp41} proposed a similar algorithm that constructs
the ZDD representing the family of all the $s$-$t$ paths.
Kawahara et al.~\cite{Kawahara2017} generalized their algorithms to
a framework that can treat various objects including matchings,
regular graphs, and Steiner trees,
and that can impose constraints such as the degree of each vertex, the connectivity of vertices, the existence of a cycle, and the number of edges (equal to, less than, or more than a specified value) in any combination.
Recent research for the construction of ZDDs enables us to
treat more complex graph classes such as degree constrained graphs~\cite{Kawahara2016},
chordal graphs~\cite{Kawahara2019}, interval graphs~\cite{Kawahara2019},
and planar graphs~\cite{Nakahata2020}. All of them can be treated as
reconfiguration objects and are shown in Fig.~\ref{fig:problems}. For example, the proposed algorithms can solve
the Steiner tree reconfiguration~\cite{DBLP:journals/ieicet/MizutaIZ17}, planar subgraph reconfiguration,
and so on.

The sliding model for edge sets can be considered by just changing the definition of the neighbor $N$ into $N(e) = \{ e' \mid e \text{ and } e' \text{ share an endpoint}  \}$.

\section{Experimental results}\label{sec:experiments} To check the performance of the proposed ZDD-based method,
we conducted an experimental comparison using the 1st place solver winning CoRe Challenge 2022. 

\noindent
\textbf{CoRe Challenge 2022.}
In 2022, the 1st Combinatorial Reconfiguration Challenge (CoRe Challenge 2022)\footnote{\url{https://core-challenge.github.io/2022/}} was held. 
CoRe Challenge 2022 is aiming for practically exploring the combinatorial reconfiguration. 
This 1st competition targets the token jumping model of the independent set reconfiguration problem~\cite{KaminskiMM12} and provides 369 instances including both instances that have a reconfiguration sequence (323 instances) and do not have any sequence (46 instances). 
Participated solvers include the state-of-the-art AI planner and BMC solvers based methods~\cite{DBLP:journals/corr/abs-2208-02495}, e.g., \textsf{SymK}~\cite{DBLP:conf/aaai/SpeckMN20} and \textsf{NuSMV}~\cite{CAV02}.

\noindent
\textbf{Experimental Condition.}
We compare our system with \textsf{recon} (@telematik-tuhh)~\cite{recon} which is based on a hybridization of the IDA* algorithm and the breadth-first search. It found 280 shortest reconfiguration sequences and became the 1st place solver of the overall solver track of the shortest reconfiguration metric\footnote{\url{https://core-challenge.github.io/2022result/}}. Except for \texttt{queen} benchmark series, instances solved by \textsf{recon} contain all instances solved by the 2nd and the 3rd place solvers. 
We compared how many shortest reconfiguration sequences can be found by each system.
We use a machine that equips a 2.30 GHz CPU and 2 TB RAM. 
\KJ{}{The proposed ZDD solver is written in C++ language with the SAPPOROBDD\footnote{\url{https://github.com/Shin-ichi-Minato/SAPPOROBDD}} and TdZdd~\cite{iwashita2013efficient} libraries and compiled by \texttt{g++} with \texttt{-O3} option.
The variable of ZDDs is ordered by the heuristic described in \cite{Fifield2020}.
The implementation of the proposed algorithm
and the scripts that reproduce the experiments are published at the GitHub page\footnote{\url{https://github.com/junkawahara/ddreconf-experiments2023}}.}

\noindent
\textbf{Results.}
Table~\ref{tab:exp1} shows comparisons between the proposed algorithm and the 1st place solver \textsf{recon} for each series of the benchmark instances. 
\KJ{}{In this experiment, the time limit is two hours.}
The 1st and 2nd columns denote the name of the benchmark series and the number of instances included in each series, respectively.
The 3rd and 4th columns denote the maximum number of vertices and edges in each series, respectively.
The longest shortest reconfiguration length known in each series is shown in the 5th column.
The 6th and 7th columns denote the number of instances solved by the two methods. 
Benchmark series are sorted in the order of the longest shortest reconfiguration length.

From this table, we can read \KJ{there are}{}pros/cons of the proposed algorithm and \textsf{recon}. The proposed algorithm is good at solving the instances having long reconfiguration sequences but not so good instances of large-sized graphs. The \textsf{recon} solver can \KJ{adopt}{adapt} instances of large-sized graphs but cannot solve \KJ{}{many of} instances having long reconfiguration sequences.
One of the major bottlenecks of the proposed algorithm is constructing the ZDD $\zind$ representing all the independent sets.
If the graph is large (e.g., the number of edges is more than 10{,}000), it takes a very long time and large memory to construct $\zind$.
Table~\ref{tab:exp2} shows the results classified by the length of the shortest reconfiguration found. This table shows more clearly the characteristics of the two methods.

\begin{figure}[tb]
  \begin{minipage}{0.49\hsize}
    \centering
  \includegraphics[width=\textwidth]{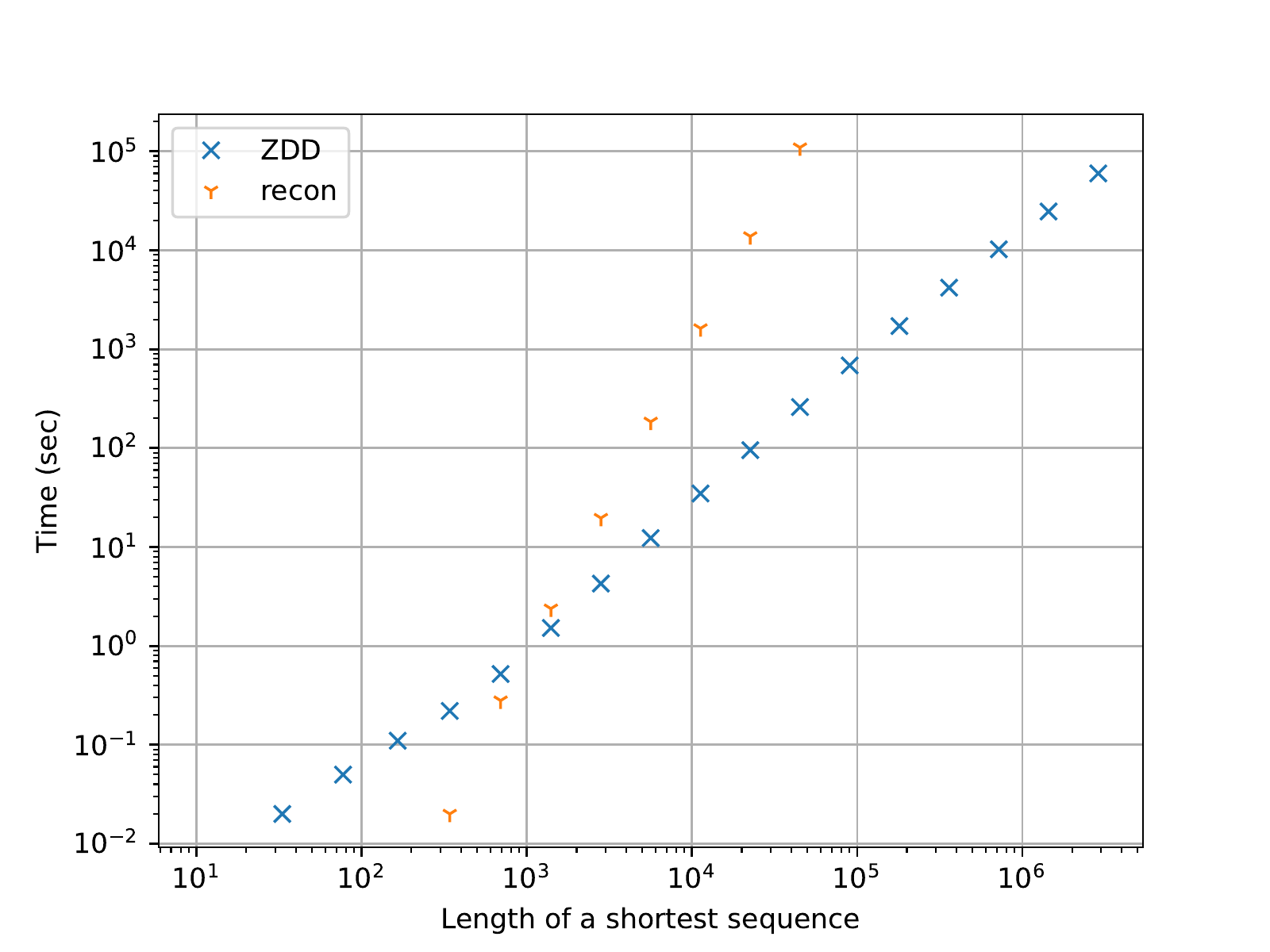}
  \caption{Time for \texttt{sp} series.}
    \label{fig:sp-cpu}
  \end{minipage}  \begin{minipage}{0.02\hsize}
  \quad
  \end{minipage}
  \begin{minipage}{0.49\hsize}
    \centering
  \includegraphics[width=\textwidth]{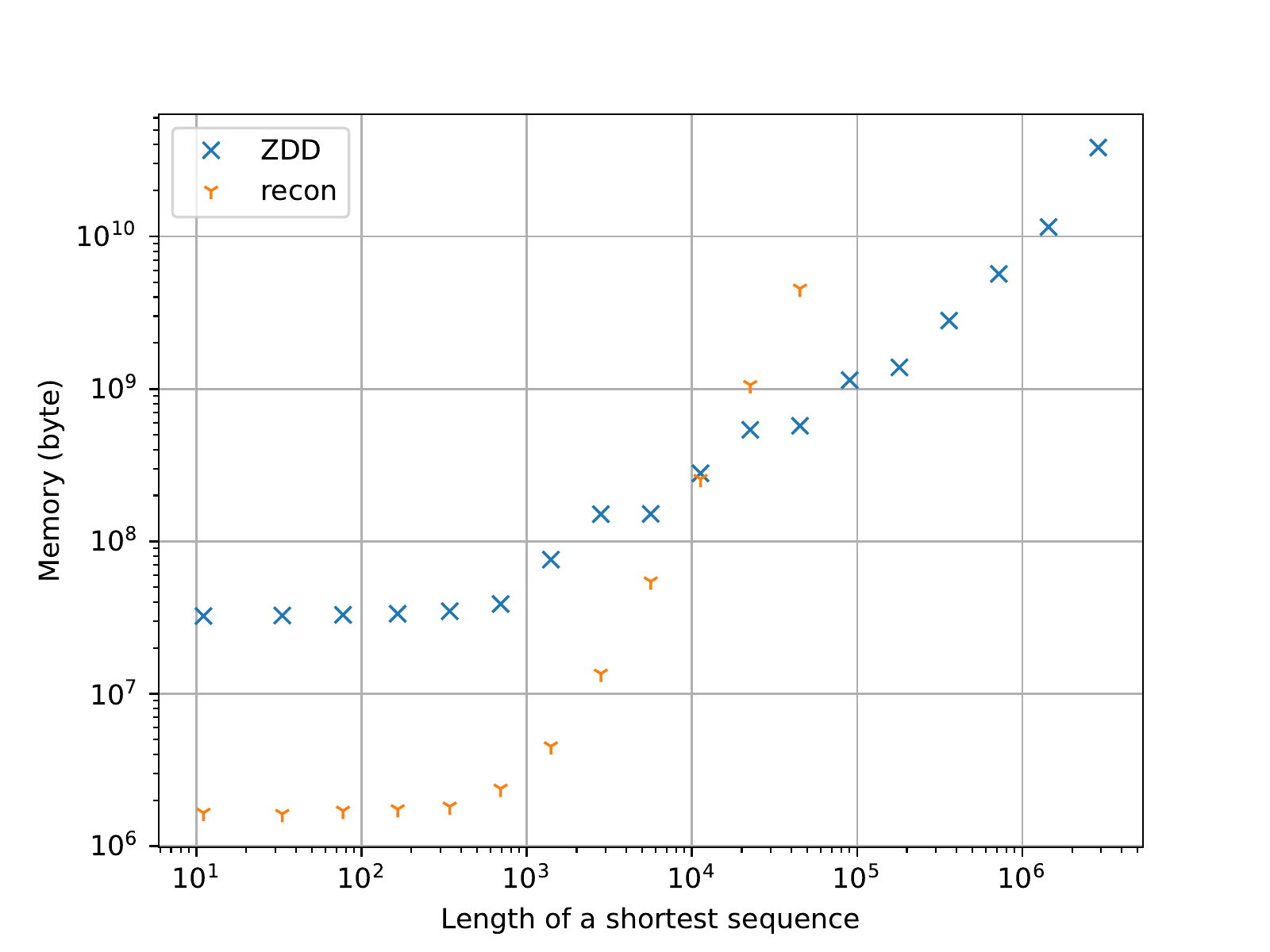}
  \caption{Memory usage for \texttt{sp} series.}
    \label{fig:sp-memory}
  \end{minipage}\end{figure}

Figure~\ref{fig:sp-cpu} shows a plot showing the relation of the length of the shortest reconfiguration sequence found and CPU time in log scale for \texttt{sp} series.
\KJ{}{In this experiment, to confirm how long the reconfiguration sequences that algorithms can find are, we set the time limit to be 200{,}000 seconds.}
We can read that the CPU time of the proposed algorithm rises gently with respect to the length, while \textsf{recon} rises steeply. Figure~\ref{fig:sp-memory} shows a similar plot, but we compare memory usage. The proposed algorithm uses the memory at the beginning for the ZDD library, but the situation reverses when the reconfiguration sequence becomes longer.
\KJ{}{The proposed algorithm succeeded in 
 computing the shortest sequence with length 5{,}767{,}157 for \texttt{sp019} instance with 247 vertices and 1{,}578 edges in 151{,}567 seconds.
In the proposed algorithm for \texttt{sp019}, the size (number of nodes) of the ZDDs $\Z^i$ for each $i$ is at most tens of thousands, but some of them contain more than $10^{10}$ independent sets.
The proposed algorithm behaves as if it searched the solution space like a breadth-first search
while compressing found solutions.
This indicates that the proposed algorithm is good at instances where the width of the solution space is relatively small, but the length of it is very long.
The \textsf{recon} solver also conducts the breadth-first search, but it takes a long time because the solution space itself is enormous.
}

\begin{table}
\caption{Experimental results for each series}
\label{tab:exp1}
\centering
\begin{tabular}{|l|r|r|r|r|r|r|} \hline
Dataset & \# Inst. & Max $|V|$ & Max $|E|$ & Reconf len & \# Solved ZDD & \# Solved recon \\ \hline \hline
\texttt{grid} & 4 & 40000 & 79600 & 8 & \textbf{2} & \textbf{2}\\ \hline
\texttt{handcrafted} & 5 & 36 & 51 & 69 & \textbf{5} & \textbf{5}\\ \hline
\texttt{color04} & 202 & 10000 & 990000 & 112 & 76 & \textbf{201}\\ \hline
\texttt{queen} & 48 & 40000 & 13253400 & 159 & 8 & \textbf{40}\\ \hline
\texttt{square} & 17 & 204 & 303 & 1722 & \textbf{17} & 8\\ \hline
\texttt{power} & 17 & 304 & 463 & 55139 & \textbf{11} & 6\\ \hline
\texttt{sp} & 30 & 390 & 2502 & 90101 & \textbf{15} & 10\\ \hline
\end{tabular}
\end{table}

\begin{table}
\caption{Experimental results for each range of reconfiguration lengths}
\label{tab:exp2}
\centering
\begin{tabular}{|r|r|r|r|} \hline
Reconf len $\ell$ & \# Instances & \# Solved ZDD & \# Solved recon \\ \hline \hline
$1 \le \ell \le 10$ & 178 & 70 & \textbf{178}\\ \hline
$10 < \ell \le 100$ & 78 & 25 & \textbf{73}\\ \hline
$100 < \ell \le 1000$ & 21 & \textbf{16} & 13\\ \hline
$1000 < \ell \le 10000$ & 11 & \textbf{11} & 6\\ \hline
$10000 < \ell \le 100000$ & 7 & \textbf{7} & 2\\ \hline
\end{tabular}
\end{table}

\section{Conclusion}\label{sec:conclusion}

We have proposed a ZDD-based framework for solving combinatorial reconfiguration problems.
We have shown that the framework can treat various reconfiguration objects, described in Fig.~\ref{fig:problems}.
The framework can solve the TAR and jumping models and the sliding model on a (directed) graph.
We have also shown that our framework is used for analyzing the solution space of reconfiguration problems such as the reachability, shortest, farthest, connectivity, and optimization variants.
We are trying to implement all of the features described in the paper,
and some of them have already been published at the website\footnote{\url{https://github.com/junkawahara/ddreconf}}.
Currently, the program can treat independent/dominating sets, matchings, (spanning/Steiner) trees, and forests as reconfiguration objects under the TAR, token jumping, and token sliding models.
We hope that these features will contribute to analyzing reconfiguration problems from theoretical and practical points of view.

Future work includes theoretically analyzing the complexity of the proposed algorithm,
designing ZDD-based algorithms for problems for which the solution space cannot be directly represented as a set family,
such as coloring reconfiguration problems~\cite{Cereceda2008} and graph partition reconfiguration problems~\cite{https://doi.org/10.1002/jgt.22856},
and applying the algorithm to practical problems.

\clearpage

\bibliographystyle{plain}
\bibliography{reference}

\end{document}